\documentclass[twocolumn,showpacs,preprintnumbers,amsmath,amssymb,aps]{revtex4}
\usepackage{bm}
\usepackage{latexsym}
\usepackage{eufrak}
\usepackage[latin1]{inputenc}
\usepackage{latexsym}
\usepackage{amsmath}
\usepackage{epsfig}
\usepackage{ifpdf}
\usepackage{graphicx}
\usepackage{dcolumn}
\usepackage{amsmath}
\usepackage{amsfonts}
\usepackage{amssymb}
\usepackage{color}

\begin{document}
\title{Braneworld Model of Dark Matter: Structure Formation}

\author{Miguel A. Garc\'{\i}a-Aspeitia\footnote{Part of the Instituto Avanzado de Cosmolog\'ia (IAC) collaboration http://www.iac.edu.mx/}$^{\circ}$} 
\email{agarca@fis.cinvestav.mx}
\author{Juan A. Maga\~na$^{*\diamondsuit}$}
%\email{jmagana@astroscu.unam.mx}
\author{Tonatiuh Matos$^{*\circ}$}
%\email{tmatos@fis.cinvestav.mx}
\affiliation{$^\circ$Departamento de F\'{\i}sica, Centro de Investigaci\'on y de Estudios Avanzados del I.P.N. Apdo. Post. 14-740 07000, D.F., M\'exico}
\affiliation{$^\diamondsuit$Instituto de Astronom\'ia, Universidad Nacional Aut\'onoma de M\'exico, A.P. 70-542, C.P. 04510, D.F., M\'exico}

\begin{abstract}
Following a previous work [Gen. Rel. Grav. \textbf{43} (2011)], we further study the behavior of a real scalar field in a hidden brane in a configuration of two branes embedded in a five dimensional bulk.
We find an expression for the equation of state for this scalar field in the visible brane in terms of the
fields of the hidden one. Additionally, we investigated 
the perturbations produced by this scalar field in the visible brane with the aim to study their dynamical properties.
Our results show that if the kinetic energy of the scalar field dominates during the early universe the perturbed scalar field could mimic the observed dynamics for the dark matter in the standard paradigm. Thus, the scalar field
dark matter hypothesis in the context of braneworld theory could be an interesting alternative to the
nature of dark matter in the Universe.
\end{abstract}

\date{\today}
\maketitle

\section{Introduction.} \label{sec_I}

Several cosmological observations \cite{WMAP} indicate the existence of matter in the Universe whose nature 
and dynamics are not predicted by the standard model of particles nor by the general theory of relativity. One component of it is the responsible of the large-scale structure 
formation in the Universe and it is called dark matter (DM) and the other component is called dark energy (DE) that could be responsible of
the late-time accelerated expansion of the Universe. Several ideas have been proposed to explain the
nature of these dark components, however, the current paradigms are still not completely satisfactory.
Thus, in order to understand the true nature of the dark components 
it is necessary to put forward alternative theories on DM and DE.

One new and promising idea is the braneworld theory that assumes the nature of DM and DE
as an effect of extra dimensions in our four dimensional Universe \cite{HW}-\cite{DGP}. 
In a previous work we supposed that the DM lives in an extra dimension 
of the Universe in a \emph{hidden brane}, and their interactions with
the fields in the \emph{visible brane} are only through gravity. 
Thus, we can detect DM by their gravitational effects but
not directly \cite{Miguel}.

With these ideas in mind, we focus on the study of the scalar perturbations 
produced by the scalar field in the hidden brane with the aim to analyze 
this kind of braneworld model as alternative
to the dark matter nature. In order to do this, we separate this study in two steps.

\begin{figure}[htp]
\centering
\includegraphics[width=8cm]{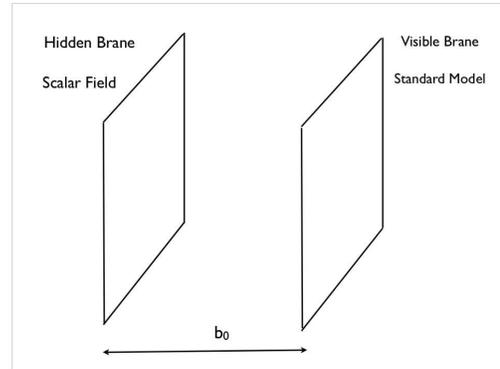}
\caption{Sketch of the model proposed with two branes: A visible brane and a hidden brane. The visible brane contents 
the standard model of particles and the hidden brane contents a real massive scalar field.} \label{fig:1}
\end{figure}

For the first step we follow \cite{Miguel}, 
where it is assumed the existence of two concentric spherical branes embedded 
in a 5D bulk to explain the DM in the Universe.
However, in the present work we will only consider a \emph{patch} of these spherical branes to obtain two local 
flat 3D branes embedded in a 5D bulk. The first brane is the \emph{hidden brane} filled only with 
a real massive scalar field, whose gravitational interaction with the 
visible brane will act as DM, the other one separated a distance $b_0$ is the \emph{visible brane} 
(our Universe) filled with the standard model of particles (See Fig. \ref{fig:1}). 
Thus, the interaction between the two branes is only gravitationaly due to the potential
wells produced by the fields residing in the branes and 
the particular topology considered.

Recent works (see for example \cite{Matos}) show that a real scalar field 
with the scalar potential $V(\Phi)=m_{\Phi}^{2}\Phi^{2}/2$ is a plausible candidate to DM.  
However, this model requires an ultralight scalar field mass of  about $m_{\Phi}\simeq10^{-22}$eV for the scalar field
 to fit the cosmological observations. 
This alternative model has been called scalar field dark matter (SFDM) 
or Bose-Einstein Condensate dark matter (BEC-DM) \cite{Urena, Urena2, lee, lee2, Ivan}, 
and it is close related with fuzzy dark matter (FDM) \cite{hu}. 
(\cite{harko}, \cite{argelia}, \cite{miguel}, \cite{further}, \cite{woo}).
Thus, in the first step we study the effects on the
visible brane induced by the scalar field of the hidden brane 
as well as the constraints due to the fifth dimensional topology.

In the second step we study the 
perturbations produced by the scalar field in the hidden brane 
and how it evolves in the visible one. In order to do that, we consider the following:

\begin{enumerate}

\item The massive scalar field in the hidden brane produces perturbations 
that influence the dynamics of the visible brane through the gravitational force.
 
\item For simplicity, we do not consider the gravitational feedback caused by the visible brane, being the only contribution to the hidden brane dynamics the scalar field. Thus, the hidden brane can be considered as isolated. 
It is important to remind that $\mathbb{Z}_{2}$ symmetry is imposed in both sides of the hidden brane.

\item The growth of the perturbations of this scalar field could mimic the dark matter behavior observed in the visible brane if they become dominant in some moment of the evolution of the Universe.

\end{enumerate}

With this scenario we study the scalar perturbations in the brane
using the modified Einstein equations (see Appendix \ref{AppendixB}).

In order to perform this study, we organize this paper as follows:  In section \ref{sec_II}  
we study the constraints in the scalar field produced by the topology assumed in Fig. \ref{fig:1}. 
In addition, the energy density, pressure, equation of state and the gravitational potential 
of the SFDM from the visible brane are analyzed.
In section \ref{sec_IV} we study the perturbations in the hidden brane produced by the scalar field 
assuming the topology aforementioned in the second step. In the section \ref{sec_V} we investigate the 
behavior of the scalar perturbations proposing a dynamical system which is solved numerically. Finally, 
in section \ref{sec_VII} a brief summary of our results, and conclusions are given.

In what follows, we work in units in which $c=\hbar=1$, unless explicitly written. 

\section{The constrained equations in a five dimensional Braneworld} \label{sec_II}

In this section we formulate the basic equations for the first step of the previously described model (for more details see \cite{Langlois}).

We start by writing the five dimensional action of the branes and bulk in the following way
\begin{eqnarray}
S[x^A,g_{(5)}]&=&-\frac{1}{2\kappa^{2}_{(5)}}\int d^{5}x\sqrt{-g_{(5)}}R_{(5)}\nonumber\\
&\pm&\sum_{i}\int d^{5}x \sqrt{-g_{(5)}}\pounds_{i},
\end{eqnarray}
where $g_{(5)}$ is the five dimensional metric, $\kappa_{(5)}$ is the five dimensional gravitational constant, $R_{(5)}$ is the five dimensional Ricci scalar and $\pounds_{i}$ corresponds to the scalar field Lagrangian for the visible and hidden brane. Then, the Einstein equation can be written in the following way
\begin{equation}
G_{AB}=\kappa^{2}_{(5)}(T_{AB}\vert_{bulk}+\widetilde{T}_{AB}\vert_{branes}), 
\end{equation}
where $A,B=0,1,2,3,4$. The energy-momentum tensor $T_{AB}\vert_{bulk}$ is for the bulk and 
$\widetilde{T}_{AB}\vert_{branes}=T_{AB}+{T}_{AB*}$ is the energy-momentum tensor for the branes, 
with $T_{AB}$ and ${T}_{AB*}$ being the energy-momentum tensor 
for the visible brane and the hidden brane respectively. 
Following the proposal of Binetruy \emph{et al.} \cite{Langlois}, it is possible to write a general 5D flat metric as
\begin{equation}
ds^{2}=-n^{2}(t,y)dt^{2}+a^{2}(t,y)\delta_{ij}dx^{i}dx^{j}+b^{2}(t,y)dy^{2}, \label{metric}
\end{equation} 
where $n(t,y)$, $a(t,y)$ and $b(t,y)$ are arbitrary functions and $y$ is allusive to the fifth coordinate. 
In the same way, we choose no energy-momentum tensor in the bulk $T_{AB}\vert_{bulk}=0$ and fix the energy-momentum tensor for the branes as
\begin{eqnarray}
T_{AB}&=&\frac{\delta(y)}{b}diag(-\rho,\vec{p},0),\nonumber
\\
T_{AB*}&=&\frac{\delta(y-1/2)}{b}diag(-\rho_{*},\vec{p}_{*},0),
\end{eqnarray}
where $\rho$ and $p$ are the energy density and the pressure of the visible brane and
$\rho_{*}$ and $p_{*}$ are the energy density and the pressure for the hidden brane, respectively. 
The visible brane is fixed in $y=0$ and the hidden brane in $y=1/2$ in the orbifold. 
The second derivative of $a$ satisfies the following differential equation \cite{Langlois} 
\begin{eqnarray}
a^{\prime\prime}&=&[a^{\prime}]_{0}(\delta(y)-\delta(y-1/2))\nonumber\\&+&\left([a^{\prime}]_{0}+[a^{\prime}]_{1/2}\right)(\delta(y-1/2)-1), \label{eq_1}
\end{eqnarray}
where $\prime$ indicates differentiation with respect to $y$, $[a^{\prime}]_{0}$ and $[a^{\prime}]_{1/2}$ denotes the jump of $a^{\prime}$ in both branes respectively. 
Evaluating Eq. \eqref{eq_1} in the Einstein tensor $G_{AB}$ (Appendix \ref{Appendix}), it is possible to obtain the following dynamical equations
\begin{eqnarray}
\frac{[a^{\prime}]_{0}}{a_{0}b_{0}}&=&-\frac{\kappa^{2}_{(5)}}{3}\rho, \label{eq_2}
\\
\frac{[a^{\prime}]_{1/2}}{a_{1/2}b_{1/2}}&=&-\frac{\kappa^{2}_{(5)}}{3}\rho_{*}, \label{eq_3}
\end{eqnarray}
where the subscript $0$ and $1/2$ for $a$, $b$ means that these functions are taken in $y=0$ and $y=1/2$ respectively. In the same way, we obtain $[a^{\prime}]_{0}/b_{0}=-[a^{\prime}]_{1/2}/b_{1/2}$.

The substitution of Eqs. \eqref{eq_2} and \eqref{eq_3} into the last equation, leads to the following constraint for the different energy densities in both branes
\begin{equation}
a_{0}\rho=-a_{1/2}\rho_{*}. \label{eq_5}
\end{equation}
Analogously, we find similar equations for $n$. Thus, we obtain the following second constraint \cite{Langlois}
\begin{equation}
(3p+2\rho)n_{0}=-(3p_{*}+2\rho_{*})n_{1/2}. \label{eq_6}
\end{equation}

In what follows, we use the Eqs. \eqref{eq_5} and \eqref{eq_6}
to obtain an expression for the equation of state (EoS) $\omega$ 
of the visible brane in terms of the EoS $\omega_{*}$ of the hidden brane.

Observe that the most general solution for $a(t,y)$ and $n(t,y)$ can be written as
\begin{eqnarray}
a(t,y)=a_{0}(t)f(\gamma\vert y\vert), \label{eq_7}
\\
n(t,y)=n_{0}(t)f(\mu\vert y\vert). \label{eq_8}
\end{eqnarray}
Here $f$ is an arbitrary function of $\vert y\vert$, while $\gamma$ and $\mu$ are given by \cite{Juan}
\begin{eqnarray}
\gamma=-b_{0}\kappa^{2}_{(5)}\frac{\rho_{*}}{6A(\vert y\vert)}, \label{lam}
\\
\mu=b_{0}\kappa^{2}_{(5)}\frac{\rho_{*}(2+3\omega_{*})}{6\widehat{A}(\vert y\vert)}, \label{mu}
\end{eqnarray}
where $A(\vert y\vert)=df(\gamma\vert y\vert)/d(\gamma\vert y\vert)$ and $\widehat A(\vert y\vert)=df(\mu\vert y\vert)/d(\mu\vert y\vert)$. If we use the equations \eqref{eq_7} and \eqref{eq_8} into equations \eqref{eq_5} and \eqref{eq_6} it is possible to write down $\omega$ in terms of $\omega_{*}$ as \cite{Juan}
\begin{equation}
\omega=\frac{1}{3}\left((2+3\omega_{*})\frac{f(-\gamma(2+3\omega_{*})/2)}{f(\gamma/2)}-2\right), \label{omega2}
\end{equation} 
where we have introduced $p=\omega\rho$, $p_{*}=\omega_{*}\rho_{*}$ and the relation $\mu=-\gamma(2+3\omega_{*})$.
With this last Eq. \eqref{omega2}, we can obtain the EoS 
of one component of the hidden brane in the visible brane.
Following Binetruy \emph{et. al.} \cite{Langlois}, let us consider linear solutions  
in order to apply them in the generalized $\omega$ function
\begin{eqnarray}
a(t,y)&=&a_{0}(t)\left(1+\gamma\vert y\vert\right), \label{sol1}\\
n(t,y)&=&n_{0}(t)\left(1+\mu\vert y\vert\right), \label{sol2}\\
b(t,y)&=&b_{0}, \label{sol3}
\end{eqnarray}
where $A(\vert y\vert)=\widehat{A}(\vert y\vert)=1$ for the linear solution \cite{Juan}. 
Using Eqs. \eqref{sol1} and \eqref{sol2} into Eq. \eqref{omega2} we obtain $\omega$ as
\begin{equation}
\omega=\frac{1}{3}\left( (2+3\omega_{*})\frac{2+b_{0}H_{1/2}(2+3\omega_{*})}{2-b_{0}H_{1/2}}-2\right), \label{omega3}
\end{equation}
where $\gamma=-b_{0}H_{1/2}$, $H_{1/2}$ is the Hubble parameter of the hidden brane 
(the subscript $1/2$ for $H$ indicates that the function is taken in $y=1/2$). Observe that 
Eq. \eqref{omega3} relates $\omega$ with $\omega_{*}$ and the expansion rate $H$ of the hidden brane.
Thus, the equation for $\rho$ in terms of $\rho_{*}$ is given by
\begin{equation}
\rho=-\rho_{*}(1-b_{0}H_{1/2}). \label{rho}
\end{equation}
To obtain the pressure in the visible brane we can use the relation $p=\omega\rho$. 

Let us consider a real scalar field $\Phi_{*}$ that lives in the hidden brane
endowed with the following quadratic scalar potential \cite{Matos,argelia,harko,miguel,further,Urena,Urena2,Guzman,Ivan,hu,lee,lee2,woo}
\begin{equation}
V(\Phi_{*})=\frac{1}{2}m_{\Phi*}^{2}{\Phi_{*}^{2}}, \label{potential}
\end{equation}
being $m_{\Phi*}$ the mass of this scalar field.
 With the quadratic potential \eqref{potential} it is possible to write the energy density and the pressure associated 
with this particular scalar field as
\begin{equation}
\rho_{\Phi*}=\frac{1}{2}(\dot{\Phi}_{*}^{2}+m_{\Phi*}^{2}{\Phi}_{*}^{2}), \; \; p_{\Phi_{*}}=\frac{1}{2}(\dot{\Phi}_{*}^{2}-m_{\Phi*}^{2}{\Phi}_{*}^{2}), \label{pd}
\end{equation}
where the dot denotes the derivative with respect to the proper time and the EoS $\omega_{\Phi_{*}}$ reads as
\begin{equation}
\omega_{\Phi*}=\frac{\dot{\Phi}_{*}^{2}-m_{\Phi*}^{2}{\Phi}_{*}^{2}}{\dot{\Phi}_{*}^{2}+m_{\Phi*}^{2}{\Phi}_{*}^{2}}, \label{se}
\end{equation}
related with $\rho_{\Phi*}$ and $p_{\Phi*}$ as $p_{\Phi*}=\omega_{\Phi*}\rho_{\Phi*}$. 
Additionally, the Newtonian potential $\phi_{*}$ associated with the energy density of this scalar field can be written as
\begin{equation}
\nabla^{2}\phi_{*}=2\pi G(\dot{\Phi}_{*}^{2}+m_{\Phi*}^{2}\Phi_{*}^{2}).
\end{equation}
With the help of the equations \eqref{omega3} and \eqref{rho} it is possible to derive 
some features of this scalar field in the visible brane.
For instance, substituting Eq. \eqref{se} in Eq. \eqref{omega3}, 
the EoS of the scalar field in the visible brane reads as
\begin{equation}
\omega_{\Phi}=\frac{1}{3}\left(\Gamma\frac{2-b_{0}H_{1/2}\Gamma}{2-b_{0}H_{1/2}}-2\right),
\end{equation}
where 
\begin{equation}
\Gamma=2+3\frac{\dot\Phi_{*}^{2}-m_{\Phi*}^{2}\Phi_{*}^{2}}{\dot\Phi_{*}^{2}+m_{\Phi*}^{2}\Phi_{*}^{2}},
\end{equation}
in the same way, $\rho_{\Phi}$ and $p_{\Phi}$ can be written as
\begin{eqnarray}
\rho_{\Phi}&=&-\frac{1}{2}(\dot\Phi_{*}^{2}+m_{\Phi*}^{2}\Phi^{2})(1-b_{0}H_{1/2}), \label{rhos}
\\
p_{\Phi}&=&-\frac{1}{6}\left(\Gamma\frac{2-b_{0}H_{1/2}\Gamma}{2-b_{0}H_{1/2}}-2\right)(\dot\Phi_{*}^{2}+m_{\Phi*}^{2}\Phi_{*}^{2})\times\nonumber\\&&(1-b_{0}H_{1/2}).
\end{eqnarray}
The Newtonian potential in the visible brane $\phi$ changes due to the presence of the hidden brane as
\begin{equation}
\nabla^{2}\phi=-2\pi G(\dot{\Phi}_{*}^{2}+m_{\Phi*}^{2}\Phi_{*}^{2})(1-b_{0}H_{1/2}). \label{Poisson}
\end{equation}
In the hypothetical case in which the hidden brane does not expand, that means $H_{1/2}\to0$, we can observe 
that $\omega_{\Phi}=\omega_{\Phi*}$, $\rho_{\Phi}=-\rho_{\Phi*}$, $p_{\Phi}=-p_{\Phi*}$
and $\nabla^{2}\phi=-\nabla^{2}\phi_{*}$. It is important to stress that the 
equations of the visible brane are similar to the equations of the hidden brane. 
The difference of the sign in the Laplacian, the pressure and the density is due to the $Z_{2}$-symmetry imposed in the braneworld model.
However, in general, the corrections of physical quantities (pressure, density, etc..) in the visible brane are related to the dynamics of the hidden brane and the behavior of the function $b(t,y)$ in the fifth dimension.

In the following sections, we study the perturbations produced by the scalar field in the hidden brane.

%%%%%%%%%%%%%%%%%%%%%%%%%%%%%%%%%%%%%%%%%%%%%%%%%%%%%%%%%%%%%%%%%%%%%%%%%
%%%%%%%%%%%%%%%%%%%%%%%%%%%%%%%%%%%%%%%%%%%%%%%%%%%%%%%%%%%%%%%%%%%%%%%%%
%%%%%%%%%%%%%%%%%%%%%%%%%%%%%%%%%%%%%%%%%%%%%%%%%%%%%%%%%%%%%%%%%%%%%%%%%

\section{Cosmological perturbations and conservation equations in the brane.} \label{sec_IV}

From here on, we will develop the second step of our study. 
We focus on the perturbed modified Einstein equations (see Appendix \ref{AppendixB})
of the hidden brane. The perturbations in the metric of this brane are caused by the presence 
of scalar field. We are interested in the scalar field perturbations 
originated during the inflationary times as well as the growth of these perturbations 
during the large-scale structure formation epoch in the visible brane via gravitational 
interactions with the hidden brane.
We remind that we are assuming \emph{no gravitational feedback} 
caused by the visible brane. 
Thus, we can redefine the scale factors and the Hubble parameters of the visible and hidden brane as $a_{0}, H_{0} \to a, H$ and $a_{1/2}, H_{1/2} \to \mathrm{a}, \mathrm{H}$.

Now, we derive the scalar cosmological perturbations of the modified Einstein equations \eqref{W}-\eqref{E4}. The scalar components of the perturbed metric 
in the conformal time $\tau$ (\cite{Ivan}, \cite{5Bertschinger}, \cite{10Kodama}) are given by
\begin{eqnarray}
{g}_{00}&=&-{\mathrm{a}(\tau)}^{2}(1+2\phi_{*}(\tau,\vec{x})),\\
{g}_{ij}&=&\mathrm{a}(\tau)^{2}{\delta}_{ij}(1-2\psi_{*}(\tau,\vec{x})),
\end{eqnarray}
where $\mathrm{a}(\tau)$ is the scale factor, $\phi_{*}(\tau,\vec{x})$ corresponds to the Newtonian potential and $\psi_{*}(\tau,\vec{x})$ is
 the spatial curvature perturbation. 

On the other hand, the perturbed energy-momentum tensor
of the hidden brane can be written as
\begin{eqnarray}
{T}_{0}^{0}&=&-(\rho_{*}+\delta\rho_{*}),\\
{T}_{i}^{j}&=&(p_{*}+\delta{p}_{*}){\delta}_{i}^{j}+\delta{\pi}_{*i}^{j},\\
{T}_{0}^{j}&=&(\rho_{*}+p_{*}){v}_{*i}=-{T}_{i}^{0},
\end{eqnarray}
where $\rho_{*}$ and $p_{*}$ are the non perturbed energy density and pressure respectively. Here
$\delta\rho_{*}$, $\delta{p}_{*}$ are the perturbed energy density and the perturbed pressure respectively, 
${\delta\pi}_{*i}^{j}={\delta\pi}_{*;i}^{j}-\frac{1}{3}{\delta}_{i}^{j}{\delta\pi}_{*k}^{k}$ is the trace-free anisotropic stress perturbation
while $v_{*i}$ is the four-velocity of the fluid.
The perturbed quadratic energy-momentum tensor is \cite{1Langloise}
\begin{eqnarray}
{\Pi}_{0}^{0}&=&-\frac{\rho_{*}}{12}(\rho_{*}+2\delta\rho_{*}),\\
{\Pi}_{i}^{0}&=&\frac{\rho_{*}}{6}(\rho_{*}+p_{*}){v}_{*i},\\
{\Pi}_{i}^{j}&=&\frac{\rho_{*}}{12}( ( 2p_{*}+\rho_{*}+2\left( 1+\frac{p_{*}}{\rho_{*}}\right)\delta\rho_{*}+2\delta{p_{*}}){\delta}_{i}^{j}\nonumber\\&-&\left(1+\frac{3p_{*}}{\rho_{*}} \right) \delta\pi_{*i}^{j}).
\end{eqnarray}
Finally, the perturbed Weyl tensor ${\xi}_{i}^{j}$ is
\begin{eqnarray}
-{\xi}_{0}^{0}&=&-{\kappa}_{(4)}^{2}({\rho}_{*\xi}+\delta{\rho}_{*\xi}),\\
-{\xi}_{i}^{0}&=&{\kappa}_{(4)}^{2}\delta{q}_{*\xi;i},\\
-{\xi}_{i}^{j}&=&{\kappa}_{(4)}^{2}(({p}_{*\xi}+\delta{p}_{*\xi})\delta_{i}^{j}+\delta{\pi}_{*\xi{i}}^{j}),
\end{eqnarray}
where $\delta{q}_{*\xi}=(\rho_{*\xi}+p_{*\xi})v_{*\xi}$.  
Two of the 5-dimensional Einstein equations are equivalent 
to the conservation equations, they can be written as
\begin{eqnarray}
\dot{\delta\rho}_{*}+3\frac{\dot{\mathrm{a}}}{\mathrm{a}}(\delta\rho_{*}+\delta{p}_{*})-3\dot{\psi}_{*}(\rho_{*}+p_{*})+{\nabla}^{2}{\delta{q}}_{*}=0, \label{econs}
\\
\dot{\delta{q}}_{*}+4\frac{\dot{\mathrm{a}}}{\mathrm{a}}\delta{q}_{*}+{\partial}_{i}(\delta{p}_{*})+(\delta\rho_{*}+\delta{p}_{*}){\phi}_{*i}=0,
\end{eqnarray}
where $ ^\centerdot \equiv{\frac{d}{d\tau}}$ is the conformal differentiation. 
We remind the relationship between the conformal and cosmological time given by $\frac{d}{d\tau}=\mathrm{a}\frac{d}{dt}$.
 
The energy-momentum tensor associated with the scalar field is
$
{T}_{ij}={\Phi}_{*,i}{\Phi}_{*,j}-\frac{1}{2}{g}_{ij}({g}^{\alpha\beta}{\Phi}_{*,\alpha}{\Phi}_{*,\beta}+2V(\Phi_{*})),
$
thus, if we perturb the scalar field as $\Phi_{*}(\tau,\vec{x})={\Phi}^{(0)}_{*}(\tau)+\delta\Phi_{*}(\tau,\vec{x})$, 
we obtain the perturbed energy-momentum tensor (here, the superscripts $(0)$ denote
a non perturbed quantity)
\begin{eqnarray}
\delta{T}_{0}^{0}&=&-{\mathrm{a}(\tau)}^{-2}(\dot{\Phi}_{*}^{(0)}\delta\dot{\Phi}_{*}-\phi_{*}\dot{\Phi}_{*}^{(0)2})-{V}_{,\Phi}\delta\Phi_{*}\nonumber\\&=&-\delta\rho_{*},\label{M1}
\\
\delta{T}_{i}^{j}&=&{\mathrm{a}(\tau)}^{-2}(\dot{\Phi}_{*}^{(0)}\delta\dot{\Phi}_{*}-\phi_{*}\dot{\Phi}_{*}^{(0)2}){\delta}_{i}^{j}-{V}_{,\Phi}\delta\Phi_{*}{\delta}_{i}^{j}\nonumber\\&=&\delta{p}_{*},
\\
\delta{T}_{i}^{0}&=&-{\mathrm{a}(\tau)}^{-2}(\dot{\Phi}_{*}^{(0)}\delta\Phi_{*,i}),
\end{eqnarray}
where we have set $\delta{\pi}_{*i}^{j}=0$ because for a scalar field we assume that non-local effects produced 
by the Weyl tensor are negligible. 
On the other hand, the quadratic energy-momentum tensor can be written by
\begin{eqnarray}
\delta{\Pi}_{0}^{0}&=&-\frac{1}{12}(\dot{\Phi}_{*}^{(0)2}+2\mathrm{a}^{2}{V}^{(0)})(\mathrm{a}^{-2}(\dot{\Phi}_{*}^{(0)}\delta\dot{\Phi}_{*}-\phi_{*}\dot{\Phi}_{*}^{(0)2})\nonumber\\&-&{V}_{,\Phi}\delta\Phi_{*}),\\
\delta{\Pi}_{i}^{0}&=&-\frac{1}{12}(\dot{\Phi}_{*}^{(0)2}+2\mathrm{a}^{2}{V}^{(0)})(\mathrm{a}^{-2}(\dot{\Phi}_{*}^{(0)}\delta\Phi_{*,i})),\\
\delta{\Pi}_{i}^{j}&=&\frac{1}{12}(\frac{3}{4}\dot{\Phi}_{*}^{(0)4}-\mathrm{a}^{4}{V}^{(0)2}+\mathrm{a}^{2}\dot{\Phi}_{*}^{(0)2}{V}^{(0)}+2\dot{\Phi}_{*}^{(0)2}\delta{\rho}_{*}\nonumber\\&+&(\dot{\Phi}_{*}^{(0)2}+2\mathrm{a}^{2}{V}^{(0)})\delta{p}_{*})\delta_{i}^{j},
\end{eqnarray}
therefore, the projected Weyl Tensor is
\begin{equation}
\delta{\xi}_{\mu}^{\nu}\equiv0.
\end{equation}
To derive the perturbed Klein-Gordon equation \cite{7Karim} we use the equation \eqref{econs}, we obtain
\begin{eqnarray}
\ddot{\delta\Phi}_{*}&+&2\frac{\dot{\mathrm{a}}}{\mathrm{a}}\dot{\delta\Phi}_{*}-\dot{\phi}_{*}\dot{\Phi}_{*}-3\dot{\Phi}_{*}\dot{\psi}_{*}+\mathrm{a}^{2}{V}_{,\Phi\Phi}\delta\Phi_{*}+2\mathrm{a}^{2}\phi_{*}{V}_{,\Phi}\nonumber\\&-&{\nabla}^{2}\delta\Phi_{*}=0. \label{pereq}
\end{eqnarray}
On the other hand, the perturbed Einstein field equations can be written as
\begin{equation}
{\delta{G}}_{\mu}^{\nu}+{\Lambda}_{(4)}{\delta}_{\mu}^{\nu}={\kappa}_{(4)}^{2}{\delta{T}}_{\mu}^{\nu}+{\kappa}_{(5)}^{4}{\delta{\Pi}}_{\mu}^{\nu}. \label{MF}
\end{equation}
Using the Eqs. \eqref{M1}-\eqref{MF}, the field equations can be written in the cosmological time 
(denoted by subscript zero $\Phi_{,0}$) as follows
\begin{widetext}
\begin{eqnarray}
6\mathrm{H}({\psi}_{*,0}+\mathrm{H}{\phi_{*}})-\frac{2}{\mathrm{a}^{2}}{\nabla}^{2}{\psi_{*}}-{\Lambda}_{(4)}&=&-\left( {\kappa}_{(4)}^{2}+\frac{{\kappa}_{(5)}^{4}}{12}({\Phi}_{*,0}^{(0)2}+2{V}^{(0)})\right)({\Phi}_{*,0}^{(0)}{\delta\Phi}_{*,0}-\phi_{*}{\Phi}_{*,0}^{(0)2}+{V}_{,\Phi}{\delta\Phi_{*}}), \label{per1}
\\
2{(\mathrm{H}{\phi_{*}}+{\psi}_{*,0})}_{,i}-\mathrm{a}{\Lambda}_{(4)}&=&({\kappa}_{(4)}^{2}+\frac{{\kappa}_{(5)}^{4}}{12}[{\Phi}_{*,0}^{(0)2}+2{V}^{(0)}]){\Phi}_{*,0}^{(0)}{\delta{\Phi}}_{*,i}, \label{per2}\\
-\frac{2}{3{\mathrm{a}}^{2}}{(\psi_{*}-\phi_{*})}_{i}^{j}+{\Lambda}_{(4)}{\delta}_{i}^{j}&=&{\kappa}_{(4)}^{2}{\delta{T}}_{i}^{j}+{k}_{(5)}^{4}{\delta\Pi}_{i}^{j}\hskip1cm ({i}\ne{j}), \label{per4}
\end{eqnarray}
\begin{eqnarray}
&&2\left[{\psi}_{*,00}+\mathrm{H}({\phi}_{*,0}+2{\psi}_{*,0})+(2\dot{\mathrm{H}}+\mathrm{H}^{2})\phi_{*}-\frac{1}{3{\mathrm{a}}^{2}}{\nabla}^{2}(\psi_{*}-\phi_{*})\right]+{\Lambda}_{(4)}={\kappa}_{(4)}^{2}[{\Phi}_{*,0}^{(0)}{\delta\Phi}_{*,0}-\phi_{*}{\Phi}_{*,0}^{(0)2}-{V}_{,\Phi}\delta\Phi_{*}]\nonumber\\&+&\frac{{\kappa}_{(5)}^{4}}{12}[\frac{3}{4}{\Phi}_{*,0}^{(0)4}-{V}^{(0)2}+{\Phi}_{*,0}^{(0)2}{V}^{(0)}+2{\Phi}_{*,0}^{(0)2}({\Phi}_{*,0}^{(0)}{\delta\Phi}_{*,0}-\phi_{*}{\Phi}_{*,0}^{(0)2}+{V}_{,\Phi}\delta\Phi_{*})+({\Phi}_{*,0}^{(0)2}+2{V}^{(0)})({\Phi}_{*,0}^{(0)}{\delta\Phi}_{*,0}\nonumber\\&-&\phi_{*}{\Phi}_{*,0}^{(0)2}-{V}_{,\Phi}\delta\Phi_{*})]. \label{per3} 
\end{eqnarray}
\end{widetext}
The Klein-Gordon equation \eqref{pereq} in the cosmological time reads
\begin{eqnarray}
{\delta\Phi}_{*,00}&+&2\mathrm{H}{\delta\Phi}_{*,0}-{\phi}_{*,0}{\Phi}_{*,0}-3{\Phi}_{*,0}{\psi}_{*,0}+{V}_{,\Phi\Phi}\delta\Phi_{*}\nonumber\\&&+2\phi_{*}{V}_{,\Phi}-\frac{1}{{\mathrm{a}}^{2}}{\nabla}^{2}\delta\Phi_{*}=0, \label{per5}
\end{eqnarray}
being $\mathrm{H}$ the Hubble parameter in cosmological time.

In what follows, we rewrite the perturbed equations \eqref{per1}-\eqref{per5} in the Fourier space. 
Therefore, it is necessary to define the Fourier component of $\delta\Phi(\tau,{x}^{i})$ as
\begin{equation}
\delta\Phi_{*}(\tau,{x}^{i})=\frac{1}{{(2\pi)}^{3}}\int{{d}^{3}k}\delta\Phi_{*}({k}^{i})\exp({i{k}_{i}{x}^{i}}), \label{Fouri}
\end{equation}
where ${k}^{i}$ is the comoving wave number \cite{7Karim}. Using Eq. \eqref{Fouri},
the Einstein-Klein-Gordon equations \eqref{per1}-\eqref{per5} in the Fourier space are given by
\begin{eqnarray}
-\alpha{{\Phi}_{*,0}^{(0)}}{\delta\Phi_{*}({k}^{i})}_{,0}&=&\alpha(3\mathrm{H}{\Phi}_{*,0}^{(0)}\delta\Phi_{*}({k}^{i})-\phi_{*}{\Phi}_{*,0}^{(0)2}\nonumber\\&+&{V}_{,\Phi}\delta\Phi_{*}({k}^{i}))+\frac{2}{\mathrm{a}^{2}}{k}^{2}{\psi}_{*}, \label{Per1}
\\
2{\psi}_{*,0}&=&-2\mathrm{H}\phi_{*}+\alpha{{\Phi}_{*,0}^{(0)}}\delta\Phi_{*}({k}^{i}), \label{Per2}
\end{eqnarray}
and the equation \eqref{per3}, in the following form
\begin{widetext}
\begin{eqnarray}
2[{\psi}_{*,00}&+&\mathrm{H}({\phi}_{*,0}+2{\psi}_{*,0})+(2\dot{\mathrm{H}}+\mathrm{H}^{2})\phi_{*}+\frac{1}{3\mathrm{a}^{2}}{k}^{2}(\psi_{*}-\phi_{*})]=\kappa_{(4)}^{2}\left(\Phi_{*,0}\delta\Phi_{*}(k^{i})_{,0}-\phi_{*}\Phi_{*,0}^{2}-V_{\Phi}\delta\Phi_{*}(k^{i})\right)\nonumber\\&+&\frac{1}{2}\beta\left(\frac{3}{2}{\Phi}_{*,0}^{(0)2}-{V}^{(0)}+2({\Phi}_{*,0}^{(0)}{\delta\Phi_{*}({k}^{i})}_{,0}-\phi_{*}{\Phi}_{*,0}^{(0)2})\left(2+\frac{\Phi_{*,0}^{(0)2}-2V^{(0)}}{\Phi_{*,0}^{(0)2}+2V^{(0)}}\right)-{V}_{,\Phi}\delta\Phi_{*}({k}^{i})\right), \label{Per3} 
\end{eqnarray}
\end{widetext}
where $\alpha$ and $\beta$ are defined as
\begin{eqnarray}
\alpha&\equiv&\kappa^{2}_{(4)}\left(1+\frac{\rho_{\Phi_{*}}}{\sigma}\right), \label{Per5}\\
\beta&\equiv&\left(\frac{{\kappa}_{(5)}^{4}}{12}\right)\left({\Phi}_{*,0}^{(0)2}+2{V}^{(0)}\right)=\kappa_{(4)}^{2}\frac{\rho_{\Phi_{*}}}{\sigma}. \label{Per6}
\end{eqnarray}
The equation \eqref{per4} reads
\begin{equation}
-\frac{2}{3{\mathrm{a}}^{2}}{(\psi_{*}-\phi_{*})}_{i}^{j}={\kappa}_{(5)}^{4}{\delta\Pi}_{i}^{j} \;\;\;\;  ({i}\ne{j}). \label{Per4}
\end{equation}
Finally, the Klein-Gordon equation \eqref{per5} in Fourier space can be written as
\begin{eqnarray}
{\delta\Phi_{*}({k}^{i})}_{,00}&+&2\mathrm{H}{\delta\Phi_{*}({k}^{i})}_{,0}+\left(\frac{{k}^{2}}{\mathrm{a}^{2}}+{V}_{,\Phi\Phi}\right)\delta\Phi_{*}({k}^{i})\nonumber\\&=&{\phi}_{*,0}{\Phi}_{*,0}^{(0)}+3{\Phi}_{*,0}^{(0)}{\psi}_{*,0}-2\phi_{*}{V}_{,\Phi}, \label{Per7}
\end{eqnarray}
where  for simplicity we have assumed non cosmological constant $\Lambda_{(4)}= 0$ in the hidden brane.

\section{Dynamical System with quadratic scalar potential in the hidden brane.} \label{sec_V}

In this section we numerically solve the system \eqref{Per1}-\eqref{Per7}
to show the behavior of the scalar field perturbations in the hidden brane.
In order to do this, we consider the simplest scalar potential 
for the SFDM given by the Eq. (\ref{potential}).
If we substitute it in the Eqs. \eqref{Per1}-\eqref{Per7}, the system
transforms into an autonomous dynamical system whose solutions can be obtained using a numerical code.
Before we make the calculations for the perturbed equations, 
it is important to introduce the Friedmann equation for the non perturbed system
in this numerical code in the following way
\begin{equation}
3\mathrm{H}^{2}=\kappa_{(4)}^2\rho_{\Phi_{*}}\left(1+\frac{\rho_{\Phi_{*}}}{2\sigma}\right),
\end{equation}
and the Raychaudhuri equation as
\begin{equation}
2\dot{\mathrm{H}}=-\kappa_{(4)}^{2}\left(1+\frac{\rho_{\Phi_{*}}}{\sigma}\right)\left(\rho_{\Phi_{*}}+p_{\Phi_{*}}\right),
\end{equation}
where $\sigma$ is the brane tension. Now, we define the following convenient dimensionless
variables for the non perturbed scalar field
\begin{eqnarray}
&&x\equiv\kappa_{(4)}\frac{\Phi_{*,0}^{(0)}}{\sqrt{6}\mathrm{H}}, u\equiv\kappa_{(4)}\frac{\sqrt{V(\Phi_{*})}}{\sqrt{3}\mathrm{H}}=\kappa_{(4)}\frac{m_{\Phi_{*}}\Phi_{*}^{(0)}}{\sqrt{6}\mathrm{H}},\nonumber\\&& y\equiv \kappa_{(4)}^{2}\frac{\rho_{\Phi_{*}}}{3\mathrm{H}^{2}},s\equiv\frac{m_{\Phi_{*}}}{\mathrm{H}}, \,\, \frac{\dot{\mathrm{H}}}{\mathrm{H}^{2}}\equiv-\frac{3}{2}\Pi.
\end{eqnarray}
Using the above definitions, observe that $\frac{\rho_{\Phi_{*}}}{\sigma}=\frac{2(1-y)}{y}$ and 
$\Pi=2\left(\frac{2-y}{y}\right)x^{2}$. Therefore we can obtain a dimensionless autonomous system for the non perturbed scalar field
\begin{eqnarray}
x^{\prime}&=&-3x-su+\frac{3}{2}x\Pi, \label{Non1} \\
u^{\prime}&=&sx+\frac{3}{2}u\Pi, \label{Non2} \\
y^{\prime}&=&-6x^{2}+3y\Pi, \label{Non3} \\
s^{\prime}&=&\frac{3}{2}\Pi s, \label{Non4}
\end{eqnarray}
being $\prime$ the differentiation with respect to the $e-$foldings number $N=\ln{\mathrm{a}}$, thus $\frac{d}{dt}=\mathrm{H}\frac{d}{dN}$.

On the other hand, for the perturbed equations \eqref{Per1}-\eqref{Per7}, we define the dimensionless variables
\begin{eqnarray}
&& z_{1}\equiv \sqrt{6}\kappa_{(4)}\delta\Phi_{*}, \,\, l_{1}\equiv\phi_{*}, \,\, U\equiv-\frac{\kappa_{(4)}}{\sqrt{6}}\frac{V_{,\Phi*}}{\mathrm{H}^{2}}, \nonumber\\ && x_{1}\equiv\psi_{*}, \,\, x_{2}\equiv\frac{\psi_{*,0}}{\mathrm{H}}, \,\, l_{2}\equiv\frac{\phi_{*,0}}{\mathrm{H}}, \,\, \nonumber\\ && z_{2}\equiv\sqrt{6}\kappa_{(4)}\frac{\delta\Phi_{*,0}}{\mathrm{H}}. \label{SIS1}
\end{eqnarray}
In terms of these new variables, we can obtain an autonomous dynamical system 
for the perturbed equations in the following way
\begin{eqnarray}
z_{1}^{\prime}&=&6l_{1}x-\left(3x-U\right)\frac{z_{1}}{x}-\frac{2k^{2}}{\mathrm{a}^{2}}\left(\frac{yl_{1}}{x(2-y)}\right)\frac{s^{2}}{m_{\Phi_{*}}^{2}}, \label{SIS2}
\\
l_{1}^{\prime}&=&\left(\frac{2-y}{2y}\right)xz_{1}-l_{1}, \label{SIS3}
\\
l_{2}^{\prime}&=&-l_{2}+\left(\frac{3}{2}\Pi-2\right)l_{2}+\left(3\Pi-1\right)l_{1}\nonumber\\&+&\frac{3}{2}y\left(3x^{2}-u^{2}\right)\nonumber+\frac{1}{2}(1+y)Uz_{1}\\&+&x\left(z_{2}-6l_{1}x\right)\left(\frac{1}{2}+y\left(2+\frac{x^{2}-u^{2}}{x^{2}+u^{2}}\right)\right), \label{SIS4}
\\
z_{2}^{\prime}&=&z_{2}\left(\frac{3}{2}\Pi-2\right)+12l_{1}U+24l_{2}x\nonumber\\&-&\left(\left(\frac{sk}{m_{\Phi_{*}}\mathrm{a}}\right)^{2}+1\right)z_{1}. \label{SIS5}
\end{eqnarray}
For a scalar field, the curvature perturbation and the lapse function match, 
therefore, we have that $\psi_{*}=\phi_{*}\Rightarrow x_{1}=l_{1}, x_{2}=l_{2}$. Additionally,
it is convenient to define the density parameters 
$\Omega_{\dot{\Phi}_{*}}\equiv x^2$ and $\Omega_{V}\equiv u^2$ which will be important for the numerical results.

Now, we solve numerically the non perturbed and the perturbed dynamical systems 
through a four order Runge-Kutta method. 
The initial conditions for the numerical integration for
both dynamical systems are taken much after the inflationary times 
when the scale factor is $a \sim 10^{-6}$ and 
the Hubble parameter is about $\mathrm{H}\sim10^{13}$GeV for the quadratic potential 
with an ultralight mass ($m_{\Phi}\sim10^{-32}$ GeV). 
It is worth to note that we have taken this ultralight mass instead of a massless
scalar field because in the SFDM model, this is able
to explain the cosmological observations of the density parameters of all the components of the Universe \cite{Matos}, as the rotation curves of galaxies \cite{harko} and the central density profile of 
LSB galaxies \cite{argelia}. With this mass, the critical mass of collapse for a real scalar field is just $10^{12}\,M_{\odot}$, i.e., the one observed in galactic haloes \cite{miguel}.
The central density profile of the dark matter haloes is flat \cite{argelia}. 
With this scalar field mass the amount of substructures is compatible with the observed one 
\cite{further}.

In the following, we show the numerical result for the system of equations \eqref{Non1}-\eqref{Non4} assuming two initial conditions in the density parameters $\Omega$.

\begin{figure}[htp]
\centering
\includegraphics[width=8.5cm]{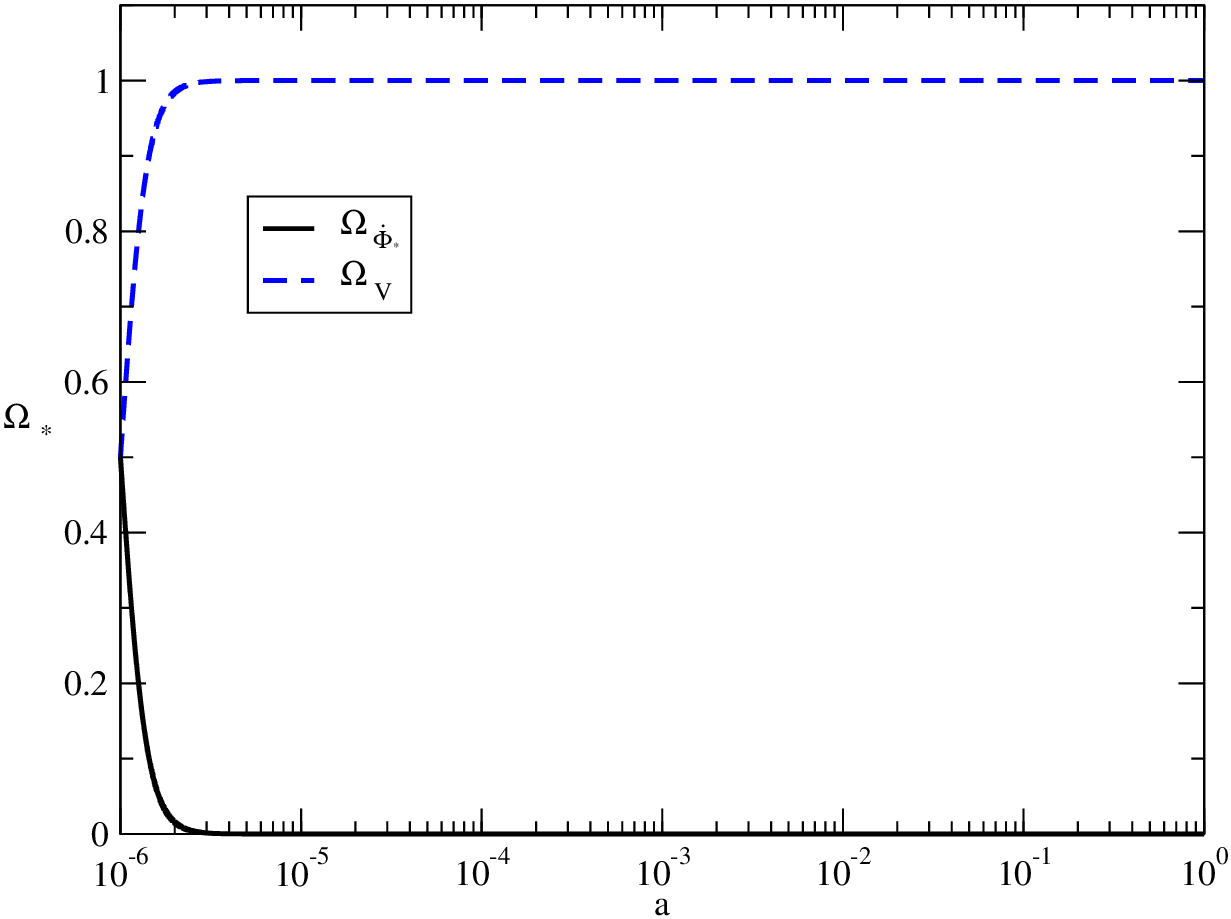}
\caption{Numerical solution for the system \eqref{Non1}-\eqref{Non4} with the initial conditions 
$\Omega_{\dot{\Phi}_{*}}\approx0.5$ and $\Omega_{V}\approx0.5$ assuming an ultralight mass $m_{\Phi}\sim10^{-32}$ GeV.} \label{fig:Nonpert1}
\end{figure}

\begin{figure}[htp]
\centering
\includegraphics[width=8.5cm]{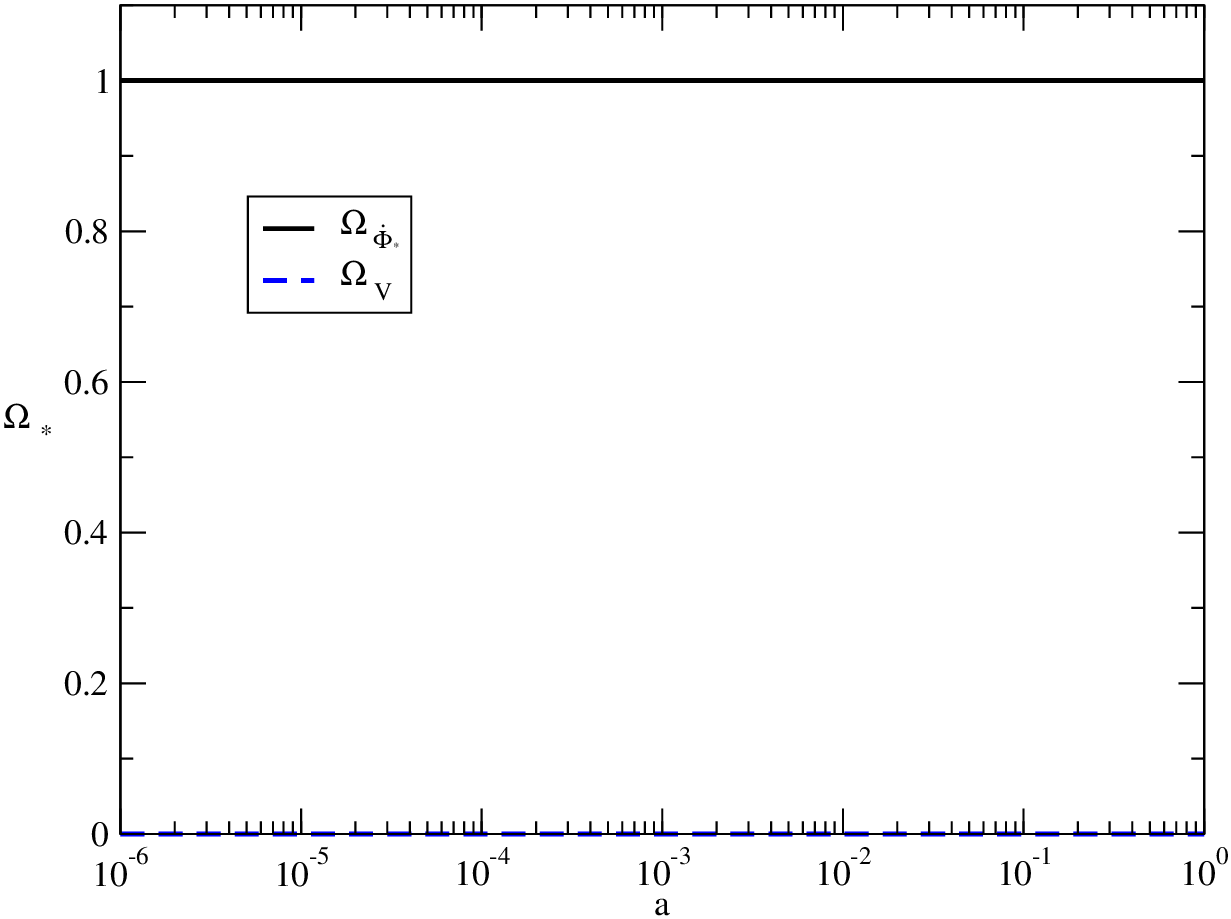}
\caption{Numerical solution for the equations \eqref{Non1}-\eqref{Non4} with the initial conditions $\Omega_{\dot{\Phi}_{*}}\approx1$ and $\Omega_{V}\approx0$ assuming an ultralight mass $m_{\Phi}\sim10^{-32}$ GeV.} \label{fig:Nonpert2}
\end{figure}

First, we take $\Omega_{\dot{\Phi}_{*}}\approx0.5$ and $\Omega_{V}\approx0.5$ as initial conditions. 
It is important to notice that we have assumed a \emph{flat metric} for the hidden brane, therefore $\Omega_{\dot{\Phi}_{*}}+\Omega_{V}=1$ at all time.
Fig. \ref{fig:Nonpert1} shows the numerical evolution for $\Omega_{\dot{\Phi}_{*}}$ and $\Omega_{V}$ with
the previous initial conditions. As we observe, the density parameter $\Omega_{\dot{\Phi}_{*}}$, 
related to the kinetic energy of SFDM, tends to zero as the hidden brane expands from early times,
at $\mathrm{a} \sim 10^{-6}$, until today at $\mathrm{a}=1$. 
On the other hand, the density parameter $\Omega_{V}$, related to
the potential energy of SFDM, becomes the dominant component ($\Omega_{V}$ $\to$ $1$) when the hidden brane expands until today ($\mathrm{a}$ $\to$ $1$).
Now, we assume that $\Omega_{\dot{\Phi}_{*}}\approx1$ and $\Omega_{V}\approx0$ 
as initial conditions, the result is shown in the Fig. \ref{fig:Nonpert2}. In this case, the kinetic energy is dominant in the evolution of the hidden brane while the potential energy is zero, 
physically this behavior is like to consider a \emph{free particle} or a kind of $k$-essence scalar field whose kinetic energy becomes dominant.

Both numerical solutions, with two different initial conditions 
($\Omega_{\dot{\Phi}_{*}}\approx0.5$, $\Omega_{V}\approx0.5$ and $\Omega_{\dot{\Phi}_{*}}\approx1$, $\Omega_{V}\approx0$), suggest different dynamics of the hidden brane filled by SFDM.
Nevertheless, the computation that will give us information about which initial condition 
causes that the scalar field behaves as dark matter in the visible brane
is the numerical evolution of the perturbed equations.

\begin{figure}[htp]
\centering
\includegraphics[width=8.5cm]{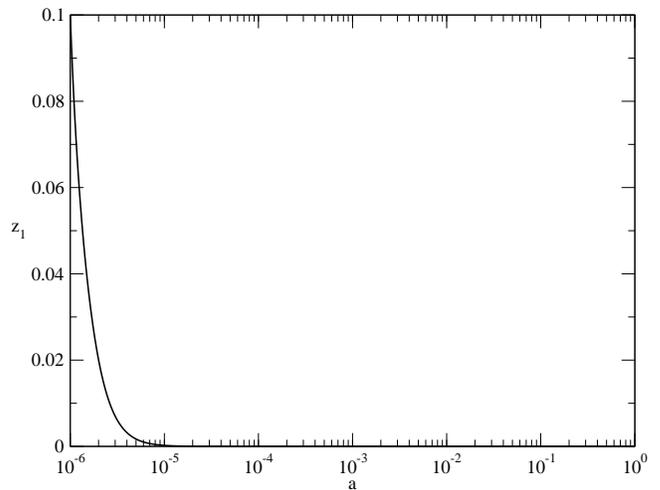}
\caption{Numerical solution for $z_{1}$ with the initial conditions $\Omega_{\dot{\Phi}_{*}}\approx0.5$ and $\Omega_{V}\approx0.5$ 
assuming an ultralight mass $m_{\Phi}\sim10^{-32}$ GeV for $k=10^{-3}$.}\label{fig:pert2}
\end{figure}

In what follows we solve numerically the dynamical system 
 \eqref{SIS2}-\eqref{SIS5} with the same ultralight scalar field.
Assuming the first initial conditions for the density parameters 
($\Omega_{\dot{\Phi}_{*}}\approx0.5$, $\Omega_{V}\approx0.5$) 
and choosing as initial conditions for the perturbations $z_{1}\approx1\times10^{-1}$, $l_{1}\approx1\times10^{-3}$, $l_{2}\approx1\times10^{-1}$, 
$z_{2}\approx1\times10^{-4}$ and $k=10^{-3}$, we obtain the numerical solutions shown in Fig. \ref{fig:pert1} and Fig. \ref{fig:pert2}.
We only plot the evolution of $z_{1}$ and $l_{1}$ because they
are directly related to the scalar perturbations and the
Newtonian potential respectively.
\begin{figure}[htp]
\centering
\includegraphics[width=8.5cm]{1-l1.eps}
\caption{Numerical solution for $l_{1}$ with the initial conditions 
$\Omega_{\dot{\Phi}_{*}}\approx0.5$ and $\Omega_{V}\approx0.5$ assuming an ultralight mass $m_{\Phi}\sim10^{-32}$ GeV
for $k=10^{-3}$.} \label{fig:pert1}
\end{figure}
Here, we observe that with these initial conditions, the scalar field does not behave 
as dark matter because its perturbation $\delta\Phi*$ and the Newtonian potential, $\phi*$, 
become dominant at very early times at $\mathrm{a} < 10^{-5}$, later on, they tend to zero at $\mathrm{a}\sim10^{-4}$,
in disagreement with observations.
It is important to stress that in the standard model, the growth of DM perturbations in the visible brane occurs in the recombination epoch at $\mathrm{a}\sim10^{-3}$. 

\begin{figure}[htp]
\centering
\includegraphics[width=8.5cm]{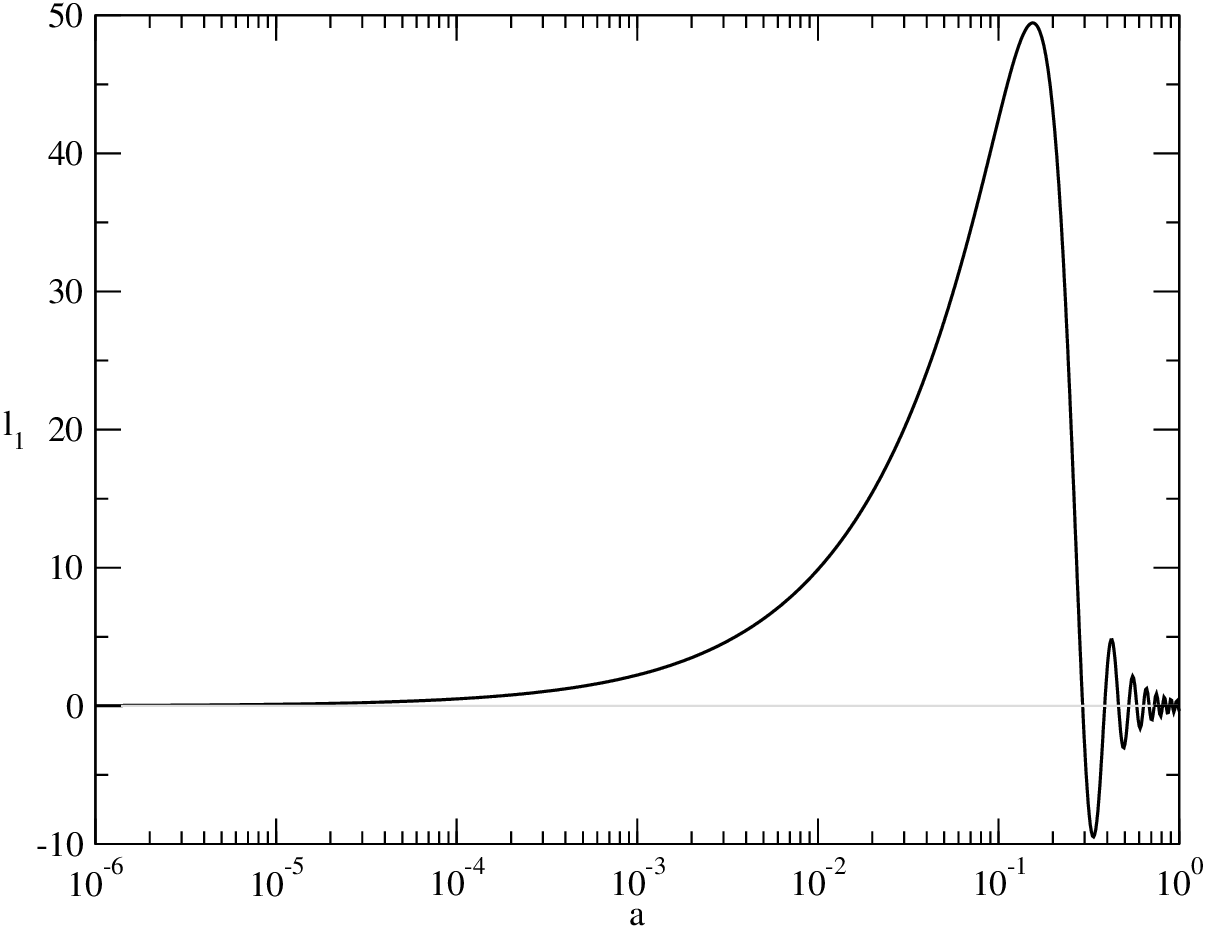}
\caption{Numerical solution for $l_{1}$ with the initial conditions $\Omega_{\dot{\Phi}_{*}}\approx1$ and $\Omega_{V}\approx0$ 
assuming an ultralight mass $m_{\Phi}\sim10^{-32}$ GeV for $k=10^{-3}$.} \label{fig:pert3}
\end{figure}

\begin{figure}[htp]
\centering
\includegraphics[width=8.5cm]{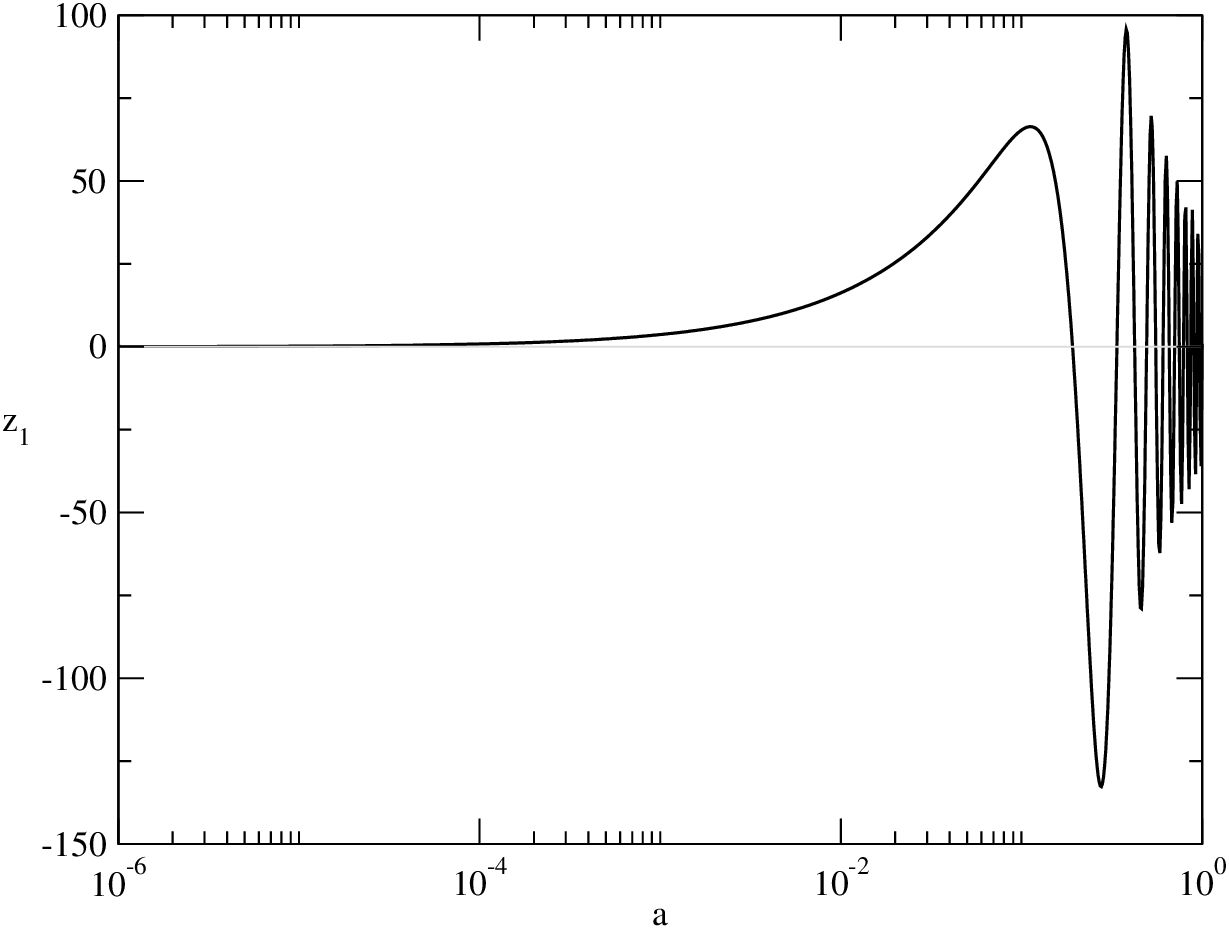}
\caption{Numerical solution for $z_{1}$ with the initial conditions $\Omega_{\dot{\Phi}_{*}}\approx1$ and $\Omega_{V}\approx0$ 
assuming an ultralight mass $m_{\Phi}\sim10^{-32}$ GeV for $k=10^{-3}$.} \label{fig:pert4}
\end{figure}

In the same way, for the case when the initial conditions for the density parameters are $\Omega_{\dot{\phi}_{*}}\approx1$ 
and $\Omega_{V}\approx0$ and choosing the same initial conditions 
$z_{1}\approx1\times10^{-1}$, $l_{1}\approx1\times10^{-3}$, $l_{2}\approx1\times10^{-1}$, $z_{2}\approx1\times10^{-4}$ and $k=10^{-3}$, we obtain the numerical solutions shown in Fig. \ref{fig:pert3} and Fig. \ref{fig:pert4}.
The observed dynamics for this case is more interesting than the observed in Figs. (\ref{fig:pert1})
and (\ref{fig:pert2}) because the induced dynamics in the visible brane is similar to
the effects of the CDM of the cosmological standard model. This is because there is a 
significantly growth in the perturbation of the scalar field, $\delta\Phi*$, 
and the Newtonian potential, $\phi*$, 
(see Fig. \ref{fig:pert3}) in the range $\mathrm{a}\sim 0.001$ 
(epoch of the domination of DM in the visible brane) to $\mathrm{a}\sim1$.
Observe that $z_{1}$, 
related to the scalar perturbation $\delta\Phi*$,
initially has a value about $\approx1\times10^{-1}$, later on,  
it reaches a maximum value  $z_{1} \sim 70$ at $a \sim 10^{-1}$.
A similar behavior is observed for $l_{1}$ that is related to the Newtonian potential $\phi*$.
The maximum of the Newtonian potential at $a\sim 10^{-1}$ (that means a redshift $z\sim 10$)
could induce gravitational wells in the visible brane that attract 
the baryonic matter to form the observed galaxies in the Universe.
Therefore, if during the early times of the hidden brane, 
the kinetic energy of the scalar field dominates, 
the scalar field fluctuations in the hidden brane would behave as dark matter in our visible brane and then, they could explain the large-scale structure formation of the universe.
This last result suggests that a scalar field in a hidden brane could mimic 
the gravitational effects of the dark matter of the standard model, 
even with an ultralight mass, this is an important result of this work. 
It is important to stress that the domain of the scalar kinetic energy could associate 
the SFDM with a k-essence scalar field as a DM source. 
However, we have to do an exhaustive exploration of the range of 
initial conditions for both no perturbed and perturbed fields. 
It is important to remark that we must study the dynamics of these braneworld models
with higher masses, which require the imposition of a scalar potential, in order
to see if this is able to recover the same results obtained in the present work.

\section{Discussion} \label{sec_VII}

In this work we study the scalar field dark matter hypothesis with an ultralight mass, $m_{\Phi}\sim10^{-32}$GeV, in the context of braneworld models using the hypothesis of Ref. \cite{Miguel}.
The conclusions and remarks can be summarized as follows.

We start assuming the topology sketched in the Fig. \ref{fig:1}, where we have 
considered a hidden brane containing a scalar field and a visible one (our universe) containing the matter of the standard model.
We obtained the modified equation of state, density, pressure and 
gravitational potential for the scalar field in the visible brane. These modifications
depend on the extra dimension and the expansion rate of the hidden brane. 
The presence of the scalar field in the hidden brane generates a new dynamics in the visible brane 
which can be considered in subsequent studies of structure 
formation in the universe or rotation curves in galaxies \cite{MAIR}.

Further, we assumed the topology aforementioned in section \ref{sec_IV}, that means, 
that there is no gravitational feedback caused by the visible brane, 
then the only contribution to the hidden brane dynamics 
will be produced by the 5D interactions and the scalar field.
 
We derived the equations of the scalar cosmological perturbations and 
projected them into the visible brane by gravitational interactions. 
We found that if the kinetic energy of the scalar field dominates 
during the evolution of the hidden brane, 
these perturbations likely behave as dark matter (SFDM) in the visible brane. 
This particular model points out that the potential associated 
with the scalar field must be subdominant and only the kinetic term will play an important role
in the dynamics of the scalar perturbations. We found a set of initial conditions
that lead to the growth of the scalar fluctuations in the hidden brane 
in the range $a\sim0.001-1$ just as in the case of the standard model.
We conjecture that these scalar perturbations generate potential wells whose gravitational 
interactions with the visible brane lead to the large-scale structure formation.
Also, our results suggest that the scalar field in the hidden brane behaves as k-essence, 
implying that we do not require a particular scalar potential.

Finally, we conclude that the braneworld model is an interesting alternative to study new dynamics 
in the Universe and it is a good alternative to explain the problem of dark matter. 
The hypothesis studied in this work gives a novel explanation to the dark matter nature
assuming two intriguing models (see for example SFDM \cite{Matos}-\cite{Ivan} and braneworld \cite{Langlois}-\cite{3Roy}). 
The results obtained in this work describe the dynamics of the SFDM and their perturbations
from the braneworld perspective. However, better results need a deeper analysis in 
the study of the formation of gravitational structure and cosmic microwave background radiation 
to obtain observational evidence of the existence of extra dimensions.

\section{acknowledgment.}

We want to acknowledge the helpful discussion with Pablo Rodr\'{\i}guez, Ivan Rodr\'{\i}guez-Montoya, Abdel P\'erez-Lorenzana and Rub\'en Cordero. 
This work was partially supported by CONACyT M\'exico, under grants 49865-F, 216536/219673 and by grant number I0101/131/07 C-234/07. J.~M. gratefully acknowledges financial support from CONACyT project 60526-F. Instituto Avanzado de Cosmolog\'ia (IAC) collaboration http://www.iac.edu.mx/.

\appendix

\section{Five Dimensional Einstein Equations} \label{Appendix}

The 5D Einstein tensor $G_{AB}$ can be obtained using the five dimensional line element (equation \eqref{metric}) 
shown in section II. 

\begin{equation}
G_{00}=3\left\lbrace\frac{\dot{a}}{a}\left( \frac{\dot{a}}{a}+ \frac{\dot{b}}{b}\right)-\frac{n^{2}}{b^{2}}\left(\frac{a^{\prime\prime}}{a}+\frac{a^{\prime}}{a}\left(\frac{a^{\prime}}{a}-\frac{b^{\prime}}{b}\right)\right)\right\rbrace,
\end{equation}

\begin{eqnarray}
G_{ij}&=&\frac{a^{2}}{b^{2}}\delta_{ij}\left\lbrace\frac{{a^{\prime}}}{a}\left( \frac{{a^{\prime}}}{a}+ 2\frac{n^{\prime}}{n}\right)\right\rbrace\nonumber\\&-&\frac{a^{2}}{b^{2}}\delta_{ij}\left\lbrace\frac{b^{\prime}}{b}\left(\frac{n^{\prime}}{n}+2\frac{a^{\prime}}{a}+2\frac{a^{\prime\prime}}{a}+\frac{n^{\prime\prime}}{n}\right)\right\rbrace\nonumber\\&+&\frac{a^{2}}{n^{2}}\delta_{ij}\left\lbrace\frac{\dot{a}}{a}\left(-\frac{\dot{a}}{a}+2\frac{\dot{n}}{n}\right)-2\frac{\ddot{a}}{a}+\frac{\dot{b}}{b}\left(-2\frac{\dot{a}}{a}+\frac{\dot{n}}{n}\right)\right\rbrace\nonumber\\&-&\delta_{ij}\frac{a^{2}\ddot{b}}{n^{2}b},
\end{eqnarray}

\begin{equation}
G_{05}=3\left(\frac{n^{\prime}}{n}\frac{\dot{a}}{a}+\frac{{a^{\prime}}}{a}\frac{\dot{b}}{b}-\frac{\dot{a}^{\prime}}{a}\right),
\end{equation}

\begin{equation}
G_{55}=3\left\lbrace\frac{a^{\prime}}{a}\left(\frac{a^{\prime}}{a}+\frac{n^{\prime}}{n}\right)-\frac{b^{2}}{n^{2}}\left(\frac{\dot{a}}{a}\left(\frac{\dot{a}}{a}-\frac{\dot{n}}{n}\right)+\frac{\ddot{a}}{a}\right)\right\rbrace,
\end{equation}
where the dots represent differentiation with respect to $t$ and the primes the differentiation with respect to $y$.

\section{The modified Einstein equations.} \label{AppendixB}

We start assuming that the five dimensional Einstein's equations are valid in the whole five dimensional space-time
\begin{equation}
{G}_{AB}^{(5)}+{\Lambda}_{(5)}{g}_{AB}^{(5)}={\kappa}_{(5)}^{2}{T}_{AB}^{(5)},
\end{equation}
where ${\kappa}_{(5)}$ is the five dimensional gravitational constant and for generalization we add ${\Lambda}_{(5)}$ as the five dimensional cosmological constant. 

In braneworlds, the energy-momentum tensor on the brane ${T}_{\mu\nu}$ and the brane tension $\sigma$
cause a discontinuity in the extrinsic curvature which is given by the Israel-Darmoise equation \cite{1Langloise}
\begin{equation}
{[{K}_{\mu\nu}]}_{-}^{+}=-{\kappa}_{(5)}^{2}\left[\frac1{3}(\sigma-T){g}_{\mu\nu}+{T}_{\mu\nu}\right], \label{ID}
\end{equation}
where $\mu,\nu=0,1,2,3$, $T={g}^{\mu\nu}{T}_{\mu\nu}$ and ${K}_{\mu\nu}={g}_{\mu}^{A}{g}_{\nu}^{B}{\bigtriangledown}_{A}{n}_{B}$,  being ${n}^{A}$ the unit normal vector to the brane and the projected metric is given by
\begin{equation}
{g}_{AB}={g}_{AB}^{(5)}-{n}_{A}{n}_{B}.
\end{equation}
In this case, we impose that the brane has a ${Z}_{2}$-Symmetry fixed in a orbifold point. 
Then, the Israel-Darmoise equation \eqref{ID} can be rewritten as
\begin{equation}
{K}_{\mu\nu}=-\frac{{\kappa}_{(5)}^{2}}{2}\left[\frac1{3}(\sigma-T){g}_{\mu\nu}+{T}_{\mu\nu}\right]. \label{IDS}
\end{equation}
On the other hand, it is possible to write the contracted Gauss equation as
\begin{eqnarray}
{R}_{\mu\nu}^{(4)}&=&{R}_{\delta\sigma}^{(5)}{g}_{\mu}^{\rho}{g}_{\nu}^{\sigma}-{R}_{\beta\gamma\delta}^{(5)\alpha}{n}_{\alpha}{g}_{\mu}^{\beta}{n}^{\gamma}{g}_{\nu}^{\delta}+{K}{K}_{\mu\nu}\nonumber\\&-&{K}_{\mu}^{\alpha}{K}_{\nu\alpha}, \label{M!}
\end{eqnarray}
and the Codacci equation
\begin{equation}
{K}_{\mu;\nu}^{\nu}-{K}_{;\mu}={R}_{\rho\sigma}^{(5)}{n}^{\sigma}{g}_{\mu}^{\rho}, \label{M2}
\end {equation}
being $";"$ the covariant differentiation with respect to ${g}_{\mu\nu}$.
The $4$-dimensional effective equations can be obtained with Eqs. \eqref{M!} and \eqref{M2}.
\begin{eqnarray}
{G}_{\mu\nu}+\frac{1}{2}\Lambda_{(5)}{g}_{\mu\nu}&=&K{K}_{\mu\nu}-{K}_{\mu}^{\sigma}{K}_{\nu\sigma}\nonumber\\&-&\frac{1}{2}{g}_{\mu\nu}({K}^{2}-{K}^{\alpha\beta}{K}_{\alpha\beta})\nonumber\\&-&{\xi}_{\mu\nu},
\end{eqnarray}
where $K={K}_{\mu}^{\mu}$ and the non local bulk gravitational field is described by 
the projected Weyl tensor on the branes \cite{1Langloise} given by
\begin{equation}
{\xi}_{\mu\nu}\equiv{C}_{AFB}^{(5)E}{n}_{E}{n}^{F}{g}_{\mu}^{A}{g}_{\nu}^{B}. \label{W}
\end{equation}
Then, using the junction conditions \eqref{IDS}, we obtain the modified Einstein equations from the view of the brane
\begin{equation}
{G}_{\mu\nu}+{\Lambda}_{(4)}{g}_{\mu\nu}={\kappa}_{(4)}^{2}{T}_{\mu\nu}+{k}_{(5)}^{4}{\Pi}_{\mu\nu}-{\xi}_{\mu\nu}, \label{E1}
\end{equation}
where
\begin{eqnarray}
{\Lambda}_{(4)}&=&\frac{1}{2}{\Lambda}_{(5)}+\frac{{\kappa}_{(5)}^{4}}{12}{\sigma}^{2}, \label{E2}
\\
{\kappa}_{(4)}^{2}&=&8{\pi}{G}_{N}=\frac{{\kappa}_{(5)}^{4}}{6}{\sigma}, \label{E3}
\\
{\Pi}_{\mu\nu}&=&-\frac{1}{4}{T}_{\mu\alpha}{T}_{\nu}^{\alpha}+\frac{1}{12}T{T}_{\mu\nu}\nonumber\\&+&\frac{1}{24}(3{T}_{\alpha\beta}{T}^{\alpha\beta}-{T}^{2}){g}_{\mu\nu}. \label{E4}
\end{eqnarray}
The conservations law equations can be obtained by ${T}_{\nu;\mu}^{\mu}=0$. 
The previous modified Einstein equations represent the behavior of the gravity in the brane with a five dimensional frame of the brane theory. 
The contributions of the brane theory to the Einstenian gravity is the quadratic part of the energy-momentum tensor \eqref{E4} and 
the non local effects produced by the Weyl tensor \eqref{W}. 
These terms will play an important role in the search of a new physics produced by the extra dimension.

\end{document}